# Remote Virtual Showdown: A Collaborative Virtual Reality Game for People with Visual Impairments

Hojun Aan, Hanyang University
Sangsun Han, Hanyang University
Hyeonkyu Kim, Hanyang University
Jimoon Kim, Hanyang University
Pilhyoun Yooh, Hanyang University
Hojun Aan, Hanyang University
Kibum Kim, Hanyang University

Many researchers have developed VR systems for people with visual impairments by using various audio feedback techniques. However, there has been much less study of collaborative VR systems in which people with visual impairments and people with able-body can participate together. Therefore, we developed a VR showdown game which is similar to a real Showdown game in which two players can play together in the same virtual environment. We incorporate auditory distance perception using the HRTF (Head Related Transform Function) based on a spatial position in VR. We developed two modes in the showdown game. One is the PVA (Player vs. Agent) mode in which people with visual impairments can play alone and the PVP (Player vs. Player) mode in which people with visual impairments can play with another player in the network environment. We conducted our user studies by comparing the performances of people with visual impairments and people with able-body. The user study results show that people with visual impairments won 67.6% of the games when competing against people with able-body. This paper reports an example of a collaborative VR system for people with visual impairments and also design guideline for developing VR systems for people with visual impairments.

## 1 INTRODUCTION

Virtual reality (VR) offers many potential opportunities for people with visual impairments, even though VR systems rely heavily on visual feedback. One of the possibilities is to play audio sports like Showdown without the showdown equipment which consists of a showdown table, a racket, and a ball. Wedoff et al. developed a virtual reality showdown game in the ball sport that is played without visual feedback [62]. In the game, the system detects the player's movement and provides spatial audio for the player to detect the location of the ball with the hearing faculty only. Wedoff et al.'s system used Kinect and Nintendo Joy-Con to detect face position and hand position. For audio feedback, researchers used HRTF (Head Related Transform Function) to allow people with visual impairments detect the location of the ball via acoustic parallax. Also, haptic feedback was provided via the controller to help people with visual impairments detect collisions between the controller and the virtual ball. So it is known that people with visual impairments can play showdown game on the VR systems. However, The VR system of Wedoff et al. is not the same as the real Showdown game because only one player can participate and a rally cannot be sustained as in a real Showdown game. Showdown is not the only system available to the visually impaired; there are also walking simulations [53, 57, 73], audio games [52], and visualization [2] VR systems for people with visual impairments.

All such VR systems for people with visual impairments are based on spatial audio feedback. HRTF is one way to provide auditory distance perception by acoustic parallax that gives each ear a different audio arrival time and audio levels depending on the location of the audio source. HRTF was used in walking simulations with collision sounds which occur between cane controller and virtual obstacle [53, 73]. It also used for environmental sounds such as virtual cars crossing a pedestrian crossing in similar walking simulation [56]. Simoês and Cavaco developed an audio VR game for people with visual impairments; their game can detect the direction of oncoming enemies using audio feedback [52]. For auditory visualization of molecular structure, people with visual impairments can understand the spatial structure of the backbone along with HRTF sound filtering [2]. HRTF allows people with visual impairments to obtain spatial information from audio feedback.

With the examples of VR systems for people with visual impairments, people with visual impairments can use VR systems, but they are still excluded from collaborative VR. Collaborative VR systems offer interactions to collaborate with other people in VR systems such as remote training and teamwork. These interactions lead to wider opportunities of collaboration in VR environment. In education, teachers can remotely teach students how to assemble an aircraft engine combustion chamber [41]. Collaborative VR can show players virtual clues via virtual hands which can point to specific components controlled by the teacher. In teamwork, two participants can design virtual robots together and learn how to design the devices that people in the real-world prefer [21]. Dollhouse VR allows designers and consumers to collaborate in a virtual house [15]. Designers can put their furniture in a top-down view, and consumers can see a virtual home in a first-person view within this VR environment. There are many opportunities like these for collaborative VR. Some researchers have extended these opportunities to people with disabilities which is a collaborative VR system for children with neurodevelopmental disabilities [31]. This VR system helps develop the emotion recognition and communication skills of children with neurodevelopmental disabilities by matching emotions with compatible therapists in VR systems.

However, because these collaborative applications rely heavily on visual information, people with visual impairments obviously have difficulties access them and their benefits. There is, therefore, a lack of research on collaborative virtual reality systems for people with visual impairments. This lack sustains the digital divide [23] in the collaborative VR systems and makes it difficult to offer possibilities for collaborative virtual reality systems to people with visual impairments.

In an attempt to provide equal collaborative VR opportunities to people with visual impairments, we have developed a multiplayer VR game in which people with visual impairments can play with other people whether they are people with visual impairments or people with able-body. We were inspired by the “Showdown” Paralympic sport which is similar to tabletop air hockey. Showdown is a sports game in which people with visual impairments and the people with able-body can play together. The reason why people with visual impairments can play the showdown game is that it relies heavily on auditory feedback. Thus, we implemented an environment using a HRTF which allows the user to detect the location of a sound source in a virtual environment. HRTF provides complex auditory information from the unpredictable movements of a fast-moving ball to people with visual impairments in a showdown game. Our game also detects and calculates about the speed at which the user swings the virtual racket and the angle at which the racket touches the ball in real time to implement a more realistic impression of ball movement.

A previous virtual showdown game used HRTF [62], but no validation study of HRTF about moving audio source was performed for the VR system's suitability for use by people with visual impairments. To provide a VR showdown gaming experience for people with visual impairments, we need to evaluate the HRTF technique and ensure that it is properly implemented in our showdown game system. Since the primary users of our collaborative VR system are people with visual impairments, we need to study the usability of our VR showdown game by people with visual impairments. To create a test system for the use of people with visual impairments and ensure that the game as a real Showdown game we developed an AI Agent that can compete against people with visual impairments. Finally, we also wanted to develop and validate a system in which people with visual impairments are able to compete against people with able-body on an equal footing. Additionally, we also test our VR sports game whether it is an exergame by measuring heart rate while gaming. Over three studies, we addressed the following research questions:

- 1. Using HRTF technology, can people with visual impairments detect the location of objects in VR environments with meaningful levels of accuracy?
- 2. Does our VR game allow people with visual impairments to compete satisfactorily with AI agents than people with able-body?
- 3. Does our collaborative VR game allow people with visual impairments to play on an equivalent level against people with able-body?

Specific research hypotheses for finding answers to the above research questions are as follows:

- Hypothesis 1. People with visual impairments will recognize the position and direction of a ball with more than 90% accuracy with auditory feedback using HRTF technology.
- In PVA (Player vs. Agent) mode, people with visual impairments have a higher probability of hitting the approaching ball into the opposing *area* and a higher probability of hitting the ball into the opponent's *goal* than people with able-body (but blind-folded).
- Hypothesis 3. In PVP (Player vs. Player) mode, people with visual impairments will defeat people with able-body more of than they will be defeated by them.

To verify the above research hypotheses, we conducted three user studies:

- User Study 1: This study measures the rate at which people with visual impairments correctly detect the position and trajectory of a showdown ball in a VR environment through auditory feedback.
- User Study 2: This study explores whether people with visual impairments can play virtual showdown games competitively with an AI agent in a virtual environment.
- User Study 3: This study explores whether people with visual impairments can play virtual showdown games competitively with people with able-body in a virtual environment.

Finally, this paper makes the following contributions:

- We have developed one of the first collaborative sports VR games in which people with visual impairments and people with able-body can play together in the same virtual environment on an equal footing.

- We conducted a spatial audio technology evaluation through a user study of people with visual impairments in the context of the showdown game environment. We confirmed that people with visual impairments can indeed track the movement of a ball with a 90 percent accuracy rate.
- We confirmed that people with visual impairments can play the VR showdown game against an AI agent competitively.
- We confirmed that people with visual impairments and people with able-body can play VR showdown games together "on a level playing field".

## 2 RELATED WORK

In this section, we introduce related works on the themes of HRTF for people with visual impairments, motion tracking for exergames, and multiplayer VR games.

### 2.1 HRTF for people with visual impairments

HRTF enables people with visual impairments to identify spatial dimensions by feeling the distance of sounds in both ears [71]. This technical characteristic has facilitated the development of various applications for people with visual impairments. These applications can be divided into three categories depending on how HRTF is used. The first type generates sound at the position in which the system responds to the user's behaviour. For example, in the VR O&M training system for people with visual impairments, a virtual cane sound is generated at the point where the cane collides with another virtual object [53, 73]. The second type generates sound at a particular location, regardless of the user's behaviour. In this type, the HRTF indicates to the people with visual impairments the location at which they need to arrive [27] or the location of the object they want to find through hearing [7]. The HRTF gives a sense of a specific location via sound, but it also has the characteristic that the sound is reflected differently depending on the characteristics of the surrounding environment. Utilizing the second characteristic, HRTF was used to implement echolocation techniques in virtual reality [1, 67]. Echolocation techniques help users understand the characteristics of their surroundings through different reflections (echoes) depending on the physical characteristics of the user's surroundings [24].

In the third type, the location where the sound is generated changes with a moving object in real-time. Thevin et al. applied the HRTF to the sound of passing cars on the streets so that people with visual impairments could recognize the passing car in a virtual environment just as they would in a real street environment [57]. Simoês and Cavaco developed an audio game in which people with visual impairments detect "aliens" using HRTF [52]. People with visual impairments try to catch aliens by turning their phones in the direction of the approaching aliens. In a recent study, HRTF was used to implement a sound generating ball so that people with visual impairments could track the movement of the ball in a virtual showdown game which is played in VR [62]. The researchers also conducted research using the characteristics that allow people with visual impairments to make mental maps using sounds generated from HRTF [43]. Using HRTF, researchers have developed applications whereby people with visual impairments can visualize visual information [14] or bimolecular structure models [2] which are difficult to describe in words. For example, "Synaestheatre," a sensory substation device with an HRTF function, can make the sound of the instrument which is playing at a distance by filtering the instrument's pre-recorded sound to match its position [14]. The system modifies the sound depending on the location of the instrument so that people with visual impairments can track the position as the instrument moves around. When rendering the molecular structure of a protein in sound, HRTF generates sound along with the spatial position of the protein's backbone [2]. The application also rendered the sound from the location of the secondary atoms by producing a different kind of sound from the previous sound. Through the above method, researchers noted that people with visual impairments could identify the spatial position of each atom in virtual reality and make a mental map of the bimolecular structure.

These studies show that HRTF can give people with visual impairments enough spatial information via audio feedback alone. However, there are not enough studies on whether people with visual impairments can fully recognize sound-generating objects, which are moving fast with acceleration. In this study, we would like to confirm that people with visual impairments can recognize and hit a showdown ball that is moving fast and complicatedly, unlike the motion of ball's whose movements, speed, and direction are controlled by the system in the showdown virtual reality games of prior research [62].

## 2.2 Motion tracking for exergames

Many exergame systems used motion tracking devices to track the user's physical movements. There are motion tracking devices that can track the user's hands or the whole body. In the past, users' hands were tracked by a Wii controller [42]. Recently, high-performance computer-based VR devices such as Vive or Oculus have emerged as VR technology has developed. For exergames, researchers used those devices to trace not only the hand movements but also the exact location of the hand in real-time. Yoo et al. tested whether state-of-the-art VR games, which demand physical movement, can produce exercise effects in players [69, 70]. They used three VR games for hand tracking: Fruit Ninia VR, HoloBall and HoloPoint. Fruit Ninia VR, and HoloBall are games that use one controller. Fruit Ninia VR aims to cut down flying fruits using a controller expressed as a sword, while HoloBall aims to hit a flying ball using a controller expressed as a racket. Holopoint is a game in which both hands are used to shoot virtual bows. Contrary to the previous example, ExerCube allows the user to hit a target directly on the wall of the virtual environment by putting a Vive tracker on the user’s wrist [33].

For tracking the whole body, previous studies used Kinect [4]. More recently, one study allowed the system to recognize the player’s walking and running movements in VR environments using Kinect [17]. However, VR-STEP allowed users to track their walking movements in smartphone VR using gyroscopes and accelerometer sensors embedded in the smartphone [58]. Qian et al. has developed a system that tracks the location of moving legs using HoloLens [45, 68]. By applying the Doppler Effect to the HoloLens connected by Wi-Fi, the technology allowed the tracking of leg movements by calculating the distance of the radio waves reflected.

Wii controllers and Kinect have also been used in exergames for people with visual impairments. VI-Bowling tracked the direction of the controller with an infrared sensor on the Wii controller to determine the direction of a bowling ball [38]. The game also tracked the speed of the user's swing with a three-axis accelerometer embedded in the Wii controller to detect the bowling action. In VI-Tennis, the system uses a three-axis accelerator on the Wii controller [37] to track the speed and angle at which the user swings a tennis racket. In Pet-N-Punch, the Wii controller tracks changes in the z-axis with an accelerator; people with visual impairments swing the controller from top to bottom mimicking the hitting of a mole with a hammer [39]. In a track and field game for the people with visual impairments, Kinect tracked the running motion of people with visual impairments, providing the feeling of running directly on an athletics track [40]. In a recent study, a “Virtual Showdown” exergame which is playable in VR, Kinect and Joy-con were used by people with visual impairments. Kinect tracked the position and orientation of the faces of people with visual impairments in order to use HRTF. Joy-con tracked changes in the hand position of the people with visual impairments on the x and z-axis position in order to calculate the angle of their hand [62].

These related studies reflect the development of motion tracking technology used in exergames. However, there remains a lack of exergames using advanced motion tracking technologies for people with visual impairments. A recent related study, Virtual Showdown, also tracked hand movements with Joy-con, but it did not fully reflect the movement of swinging controllers in the virtual environment. In the previous Virtual Showdown game, the racket is implemented to turn discretely at 45-degree intervals only when the player turn continuously [62]. By contrast, we developed a more realistic showdown game using Vive to track the position and direction of the hand in real-time.

### 2.3 Multiplayer VR game

An empirical study of previous multiplayer games was conducted. Seele et al. conducted a study on the quality of conversation in social games in virtual environments [50]. They compared and analysed cases using avatars with eyes moving based on eye-tracking technology, avatars with eyes moving based on artificial intelligence, and avatars with only eye fixation and saccadic movement. As a task of the user study, in a virtual environment where two people sit opposite each other, they introduced themselves and played a game in which they looked at the images of several people and asked about the characteristics of the person they had set up. In a questionnaire about the collaborative environment, participants answered that they were able to have high-quality communication in virtual reality regardless of the Avatar's eye movement. Christensen et al. compared the monitor/controller environment and the VR environment within the collaborative puzzle game [5]. In the game, players were to communicate and inform each other about which levers needed to be lowered. Through a questionnaire, the players confirmed that the VR environment was better in terms of immersion, concentration, empathy, behavioural involvement, positive influence, and negative influence than the monitor/controller environment.

Other studies report on serious games in collaborative VR. Wegner et al. developed a game in which two VR players were to treat patients [63]. In this game, one played the role of a senior paramedic helping another paramedic using a virtual tablet PC. The other player played the role of a junior paramedic, treating the patient directly using a virtual tool at the junior paramedic’s waist. Ha et al. developed a game in which several players put out a fire that was blocking their way so they could escape from the building together [13]. Social MatchUp is a smartphone VR game that a child with a neurodevelopmental disorder can play with a therapist [31]. The goal of the game is to find the same emoticon simultaneously; this activity aims at improving the ability to recognize and verbalize emotion.

In addition to empirical analysis, research on multiplayer VR games have been developed in various fields such as serious games and games for people with disabilities. However, there remains a lack of research regarding collaborative VR games for people with visual impairments. Therefore, we conducted a user study in which people with able-body and people with visual impairments play games together using our newly developed collaborative virtual showdown game.

## 3 VIRTUAL SHOWDOWN

### 3.1 Hardware

We used Vive Pro for multimodal feedback. Vive Pro consists of base stations, controllers, and an HMD, which includes Bluetooth communication which enables recognition of the equipment using the software programming provided by SteamVR. The dual-stage trigger feature mounted on the Vive Pro Controller is also used, and SteamVR Tracking 2.0 is used as a sensor. The Vive Pro HMD is capable of playing lossless audio which is mastered in a higher sound quality format than is commonly used. Vive HMD also has sensors such as StreamVR tracking, G-sensor, gyroscope, proximity sensor, and pupil distance setting (IPD).

*3.1.1 Motion Recognition System.* The goal of implementing the Vive HMD as a motion recognition system is to detect the user's head position, angle, and movement through Bluetooth so that the movement of the head matches the virtual head object in the VR environment. HMD’s transform was used to calculate the sound of objects (the sound of a ball rolling, the sound of a ball touching the wall, etc.), and the direction and distance from the user's head position were all used to produce realistic sounds. Fig 1 shows the field of vive tracking environment which can track HMD’s spatial movement. In the case of physical mechanics, the ball's velocity variable in Unity was implemented to achieve a virtual speed similar to the ball's speed in the original Showdown game, i.e., 30-40 km/h. To this end, the physical material setting of Unity was modified, the friction variable was set to zero, and the elastic variable of the ball was set to 0.9, applying the physical law as closely as possible to the original Showdown. In addition, the balls in our system have been implemented to keep moving only on the playing surface. This prevents errors that are made in real-life Showdown games where the ball goes out of the field of play and pauses the game. In the case of showdown rackets, the mass variable is set to eight in comparison to the weight of the actual ball which follows Unity's basic physical law system. Variables (e.g., the angle between racket and touched ball, the speed of swing) determine the speed and direction of the ball.

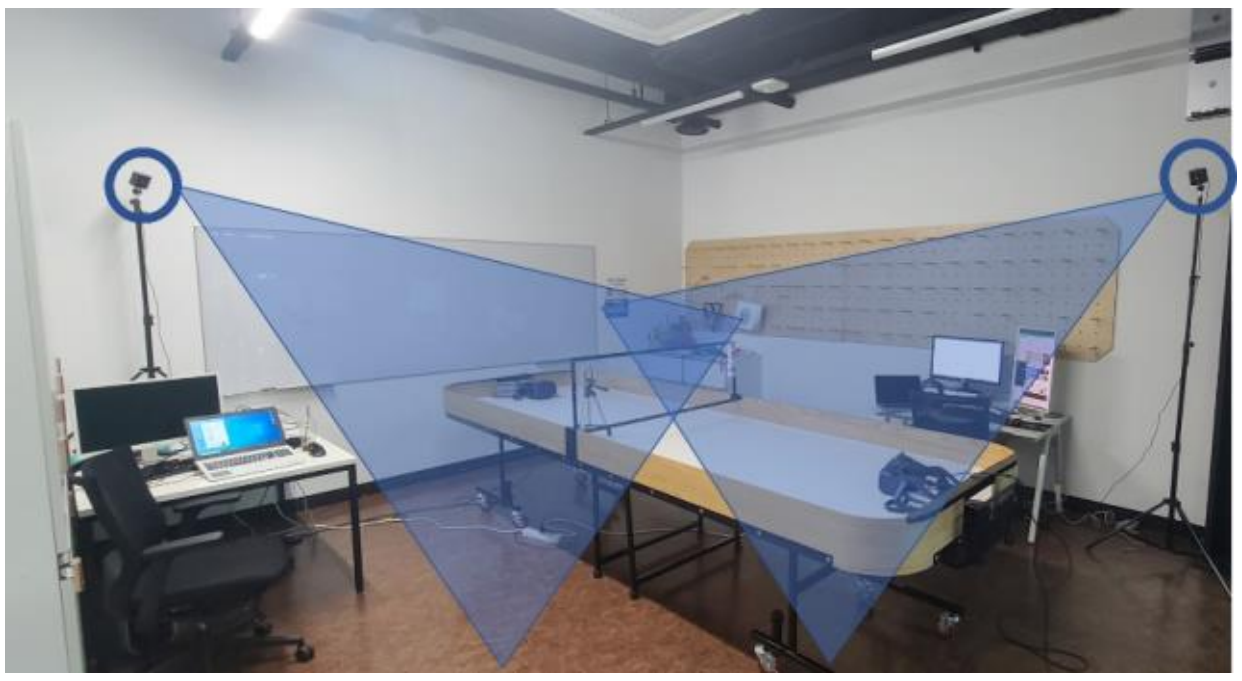

Fig 1. The Vive Tracking Environment Recognizing Entire Field

## 3.2 Sound System

In a showdown, the ball’s movement is complex. The ball may bounce against a wall or change speed and the sounds associated with such changes are complex. Since sound is the only information available to judge a situation when the eyes are covered, users need to be able to get enough information to detect the location and speed of the ball and to understand ongoing situations via hearing alone. Thus, the implementation of spatial sound is the most important part of this study.

We used HRTF technology to implement the sound system. Head-Related Transfer Function (HRTF) is a three-dimensional function technique that produces sound waves from a sound-generating object in a 3D environment in an all-round way, and then calculates the number of the sound waves and the direction in which they come to the user's head, specifically to each ear. In other words, HRTF is a technology that calculates the user's position, the centre of the head, and the distance between the right and left ear, thereby creating 3D spatiality (a sense of the spatial environment) by transmitting differences in the size and height of sounds heard in the right and left ears of the player.

HRTF is also one way to combine spatial sounds. Zahorik et al. pointed out that HRTF affects hearing distance perception [71]. The main function of the HRTF is the acoustic parallax effect, which allows the user to sense a significant difference between the direction of sound for the right ear and the left ear. They emphasized that HRTF's acoustic parallax effect explains why listeners can recognize the distance of the nearby audio source as in distance perception which is well estimated by the two eyes. Similarly, the panning system that can adjust the sound volume on the left or right side has a similar function. Thus, Larsen et al. compared the panning system and HRTF to locate and follow the virtual sound source object [29]. They found that HRTF is superior to panning systems in terms of location recognition accuracy and search speed. To detect the precise position of an audio source in virtual environment, HRTF is necessary.

*3.2.1 Google resonance sound* There are various Software Development Kits (SDKs) available for HRTF. These include Oculus Spatializer 1.22.0, Google's Resistance Audio 1.2.0, Steam Audio 2.0 Beta 13, etc. From among these, we chose to use Google resonance sound. The features of this tool include a reflect function that calculates the sound waves bouncing through the air, a direct sound function that decreases sound with increased distance, and a function that simulates environmental occlusion effects by different low and high-frequency components. Developers of this tool implemented an effect whereby sound transmission appears to come directly from the location of a virtual object through these various functions.

According to Richard Gould’s column, this tool has the advantage of having more spatial information in near-field acoustical effect compared to other tools [12]. This advantage matched the goal of our program which was to develop a system that enables users to locate a ball at close distances. In addition, the developers of Google Resonance Sound have implemented reliable real-time calculations for the implementation of spatial sound and the ability to generate sound in the correct direction of the sound source. Therefore, we chose Google resonance sound to implement HRTF because Google resonance sound lets the user accurately identify the direction of the sound source and the location of near-by sound source in real-time. Additionally, we used a standard head model provided from Google Resonance Audio SDK for the implementation of HRTF.

For the proper migration of showdown to the virtual environment, the resonance sound listener component in the Google resonance SDK was assigned to the object corresponding to the user's head position and the sound size calculation criteria of the Google reflection SDK was used without modification.

For various conditions - the sound of the ball rolling, the sound of the ball crashing into the wall, and the sound of the racket and the ball colliding - we used a recorded sound from the original Showdown game. The sound source was used for implementation in combination with the functions of HRTF. When the speed of movement of the ball drops below 10 km/h, the sound size is reduced to 9/10 of the speed of 1 km/h, and when the ball stops, no sound is heard from the ball. In addition, all continuous start and end sound wavelengths are aligned to ensure that natural sounds are produced with using repeated audio even if the sound is repeated over and over again.

## 3.3 Haptic System

In the case of vibration feedback, we implemented the software to ensure that vibrations for each situation occur on VIVE's controller when the ball touches the user's racket, when the racket hits the ball, and when the user uses the holding function. The Vive controller has an internal function called 'Pulse' which momentarily vibrates the controller. We used 'Pulse' to generate vibration frequencies of 70 per second for strong vibrations and 30 per second for weak vibrations. When the racket hits the ball, a short strong vibration is generated; while in a holding situation, where the ball continuously touches the racket for more than a second, a weak vibration is generated. This haptic feedback helps users to understand when they have hit the ball or are holding it, all without visual feedback.

### 3.4 Implementing the Competition System

From this section, we explain the basic rules and the two competition modes we implemented.

*3.4.1 Basic rules of progress for virtual showdown play* The basic rules of play from the start to end of the game follow:

- The start of a game: The sound of the ball rolling from the centre to the player signals the start of the game.
- Serve: Each player gets a serve ball that comes from the centre, twice each turn.
- Scoring the goal: Two points are scored if a ball goes into the opponent's goal pocket. After two points are scored, the game system announces (in a guiding/moderator's voice) that the player who scored has two more points. The game system starts the next serve sequence 5 seconds after the goal announcement.
- Dead ball: When the ball lies "dead" (motionless and soundless), the serve sequence is re-executed by the last player to serve and with no demerits.
- Victory: The player who scores 12 points first wins.

*3.4.2 A new function in virtual showdown* A new function, "Holding" has been added to showdown gameplay. In the original showdown game, players use the technique of "holding" the ball by gently pushing the ball to the right or left sidewall using the racket. Once the ball stops, then the player can accurately shoot the ball. In reality, friction can stop the ball's movement by pushing it against the wall with the racket. However, in the Unity engine, a series of collisions occurs between objects, walls, and rackets during such actions. If a collision between objects occurs continuously in a short time, the limit of the physical engine causes the ball to pass through the racket or to amplify the reaction of the collision, causing the ball to bounce at an abnormal speed. To address this problem, "Holding" has been implemented in our game. If the ball is in contact with the racket, and, if the player simultaneously pushes the trigger button, Holding is activated. Once holding is activated, the ball follows the racket at a constant speed. The ball's coordinates are calculated every 1/60th of a second and the ball's position is changed to follow the racket at a fixed distance from the end of the racket while the trigger button is being pressed. In the meantime, vibration feedback is provided to the controller in the hand of the user who is using the holding function. This feedback indicates whether the user is using the holding function. This will be described in more detail in the Haptic System section below.

*3.4.3 PVA (Player vs. Agent) mode* We implemented AI (Artificial Intelligence) agent for the Player vs Agent (PVA) mode where AI agent represents an opponent with visual impairment as showing in fig 2. The AI opponent was developed to participate in showdown games against real participants, so it was implemented to follow the ball like a real person and also to make mistakes while hitting the ball with a showdown racket. We developed AI to make it hard for AI to block the opponent's strong shots. The approximate criterion for shooting strong shots is set at over 11 on the velocity variable of Unity's rigid body class which contains various physical states and functions to control the states of an object. The setting of 11 here represents a threshold for shots over 38km/h in real life.

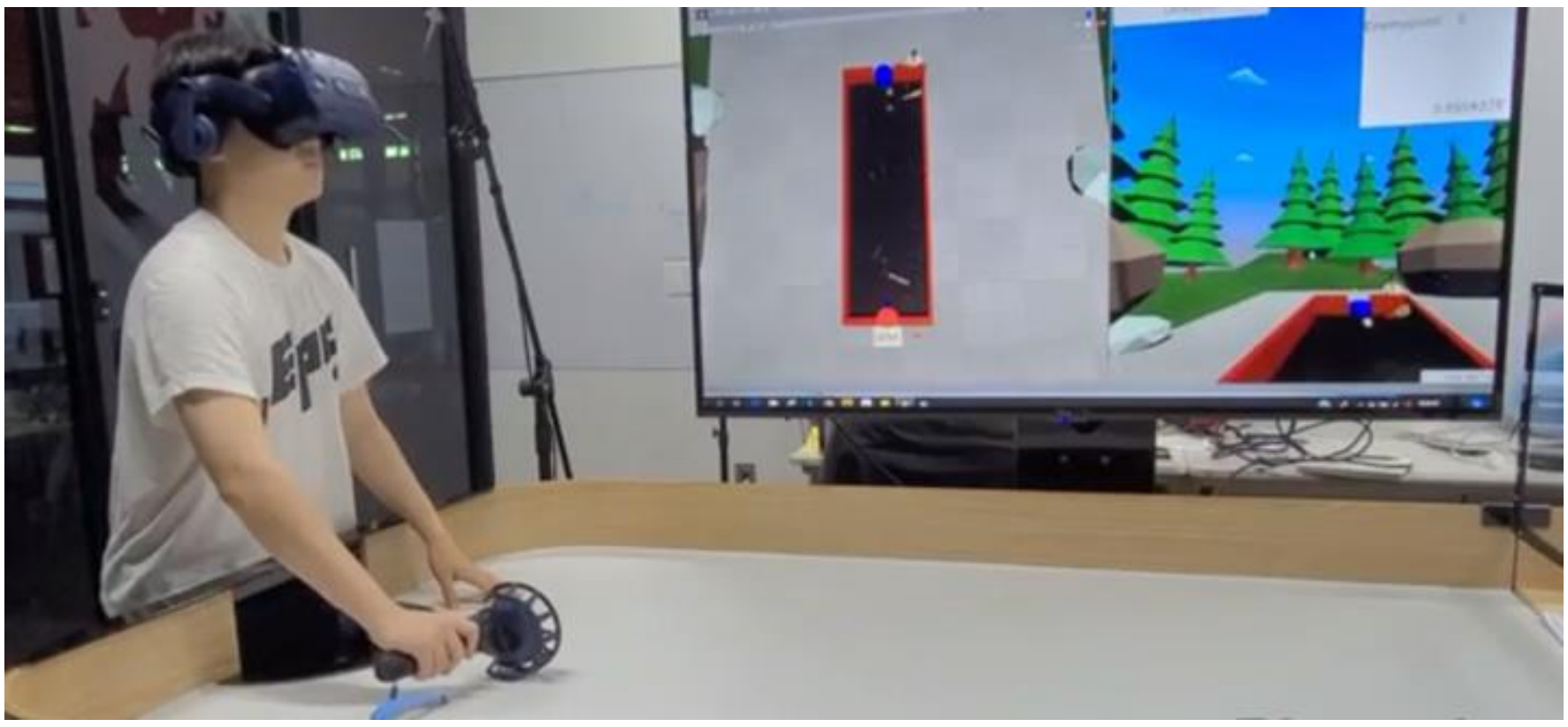

Fig 2. PVA showdown play.

Furthermore, to prevent boredom from too many rallies, the more times AI hits the ball, the less likely it becomes for AI to hit the ball. The number of cases in which AI was programmed to hit the ball is as follows:

1. In order for play to proceed beyond the serve, AI hits the approaching ball with a 100 percent success rate.
2. After hitting the ball once, the probability of AI hitting the ball reduces to 70%.
3. After hitting the ball twice, the probability of AI hitting the ball reduces to 40%.
4. AI is not programmed to precisely defend its own goal after hitting the ball three times but randomly defends either the right or the left edge of the net.

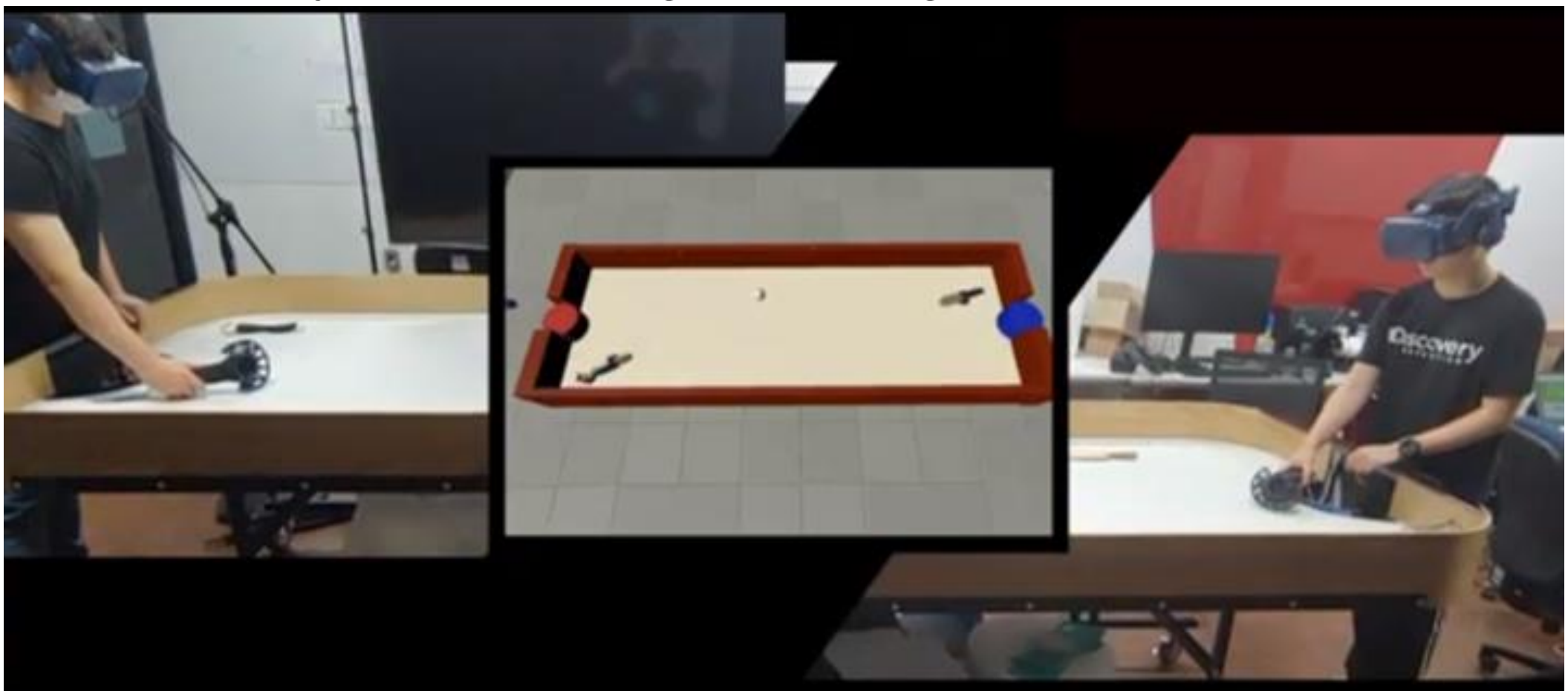

Fig 3. PVP showdown play.

*3.4.4 PVP (*Player *vs. Player) mode* For interaction between people with visual impairments and a human opponent, the multiplayer mode was developed by implementing network communication through the Photon engine from the Unity Asset Store [59] as shown in fig 3. We developed our program for each user to hear the reversed sound volume of the ball which both users listens so that each user can estimate the distance and direction to the ball with the same but mirrored sound information. The information of the ball, which is the core auditory feedback, is transmitted with the server to each computer in real-time (e.g., the position, speed, and direction of the ball). The sound of the ball is heard by each user and is calculated separately based on each user's position. Each user then hears a different sound which represents the ball’s position and trajectory relative to each player respectively. In cases where the sounds are common to both participants, a computer send same sound to another computer through server to give same sound to both user. For example, if one of the players scores a goal, the each player’s computer will play the score announcement voice through sever.

## 4 DESIGN RATIONAL

In this section, we explain why we chose specific technology or game rules from the four design elements when implementing remote virtual showdown. From virtual showdown [62], There are five elements (dimensional and physics emulation, body interaction, audio, vibration, scoring) for designing a virtual showdown game. We have implemented dimensional and physical emulation and body interaction through the HTC Vive system. We also used HRTF for the audio factor and changed the scoring system to resemble the original Showdown rules. We also added the “hold the ball” function.

### 4.1 Audio: HRTF (Head-Related Transfer Function)

HRTF was used to provide players with audio distance perception of the virtual showdown ball. Audio distance recognition is caused by binaural signals such as acoustic parallax [71]. Acoustic parallax detects significant differences in the relative angles from each ear to the virtual object audio source. HRTF is a popular way to implement acoustic parallax in a virtual environment [22, 27, 28, 53, 73]. A similar stereo audio panning system is capable of adjusting the sound level on the left or right side, however, it is not a good alternative for VR applications that require various spatial cues like acoustic parallax from auditory feedback alone as required in the showdown game. Larsen et al. compared the panning system with HRTF to locate and follow a virtual sound source object [29]. They found that HRTF is superior to the panning system in terms of location recognition accuracy and search speed.

### 4.2 Tracking system: HTC Vive

We used HTC Vive for a better player experience in VR showdown games. In a related study [62], researchers used Kinect for body tracking and Nintendo Joy-Con for hand tracking. However, Kinect does not offer as good a gaming experience for physical gameplay as Vive [32]. Besides, in virtual showdown [62], the researchers implemented the rotation of the controller in a VR system using the Z and X positions of the controller, however, it detects the angle of the controller at 45-degree intervals only. This low fidelity system can make it difficult to direct the ball precisely to the player's intended position. To overcome this problem, we used the HTC Vive Controller which can track the user's racket movement at a detailed angle i.e., with much greater precision.

### 4.3 Game rules: Very similar to the original game

We have implemented similar game rules to the original game. In the original game, players score points when they hit the ball into the opponent's goal pocket. In a related study [62] which developed a showdown arcade game, the researchers created several scoring rules, such as touching the ball, passing the ball to the opponent's area, and directing the ball into the opponent's goal pocket. However, these rules were different to the original showdown game, so original Showdown gameplay experiences are not so available. To overcome this problem we used Photon's network plugin to develop a collaborative VR game for two people to play in the same virtual environment. Since the goal of the system is to approximate to the original Showdown game as nearly as possible, we have implemented the original scoring rules, which state that the player who hits the ball into the opponent's goal pocket scores.

### 4.4 Additional interaction: Holding

While implementing our showdown game, we found that it was difficult to hold the ball. In the original Showdown game, the player holds the ball by pushing the ball to the side of the table. This strategy allows the player to move the ball to the intended launch position. However, in a VR system equipped with the Unity physics engine, it was difficult to hold the ball by pushing the ball to the sidewall because the virtual ball tried to escape from between the stick and the sidewall. Thus, we implemented a holding function that allows the player to “catch” the ball by pressing the trigger button on the VR controller. This feature is activated when the player presses the trigger button while the ball makes contact with the virtual racket. The ball follows the racket while the player is holding the trigger button. Then, when the player releases the trigger button, the ball stops at the controller's current position. This feature allows players to use the same strategy as the original Showdown game.

## 5 USER STUDY

To evaluate the virtual showdown game, we designed three different user studies. For the first user study, we tested the HRTF system in the showdown table environment to check whether Google Resonance Audio is viable for our virtual showdown application. For the second user study, we tested PVA mode (Player vs. AI Agent) to see whether people with visual impairments can play the showdown game against the AI agent. For third user study, we tested the PVP (Player vs. Player) mode to see whether people with visual impairments can play the showdown game with people with able-body. For our user study, we recruited thirteen people with visual impairments. All thirteen participants participated in all three user studies. The participants’ ages ranged from 47 to 72 (M= 61.15). All people with visual impairments were right-handed. Detailed information about the participants is shown in Table 1.

Table 1. Demographic information of people with visual impairments.

| ID | Age/Sex | Description of the Visual condition | Experience of Showdown | Experience of VR |
|---|---|---|---|---|
| V1 | 63/M | Low vision for 25 years (with light perception) | Yes | No |
| V2 | 47/M | Blind for 3 years | Yes | No |
| V3 | 58/M | Low vision for almost 30 years (with shape perception) | Yes | No |
| V4 | 74/M | Low vision for 43 years (with light perception) | No | No |
| V5 | 64/M | Low vision for 38 years (with shape perception) | Yes | Yes |
| V6 | 72/M | Blind for 60 years | No | No |
| V7 | 54/M | Low vision for 34 years (with shape perception) | Yes | No |
| V8 | 58/F | Blind for 29 years | Yes | No |
| V9 | 54/F | Low vision from birth (with shape perception) | Yes | No |
| V10 | 50/F | Low vision from birth (with shape perception) | Yes | No |
| V11 | 60/M | Low vision from birth (with shape perception), no vision perception from the left eye | Yes | No |
| V12 | 59/F | Blind for 15 years | No | No |
| V13 | 67/F | Low vision for 18 years (with light perception) | Yes | No |

### 5.1 User Study 1: Validation of HRTF in virtual showdown

In this user study, we tested whether people with visual impairments can track the spatial sound of a moving showdown ball. To validate HRTF, we hypothesized that people with visual impairments can accurately detect the start and end of ball movement to an accuracy of 90%. A related study showed that people with visual impairments detected the route of the moving ball via sound recorded by two microphones [36] to an accuracy of about 97%. We expected a similar accuracy rate in our game. However, we set our hypothesis' expected accuracy rate lower than the related study because we used the simulated binaural sound produced by HRTF instead of recorded binaural sounds.

H1. People with visual impairments will recognize the position and direction of a ball with more than 90% accuracy with auditory feedback using HRTF technology

*5.1.1 Participants and Apparatus.* 13 people with able-body and 13 people with visual impairments participated in this user study. The age range of the participants with able-body was 21~35 (M=23.76). All were university students with normal or corrected-to-normal vision and with normal hearing ability. In addition, all able-bodied participants had previous experience in VR environments. The virtual showdown software was used but participants did not move the racket in this user study. The participants only used the VR headset to listen to the moving ball. Even though the game was not played, the participants were exposed to the environment of virtual showdown.

*5.1.2 Tasks and measurement.* The participants had to detect and report the start and end positions of the moving showdown ball. Three areas immediately in front of the participant, left, middle and right, and two areas far from participant, left and right, were specified. Six departure routes - from three areas on the near side to two areas of opponent's side (3 * 2 = 6 routes) - were also specified. Six arrival routes correlated oppositely to the departure routes. We tested each route with each participant three times. Therefore, participants answered 36 questions about their sense of the trajectory of the each ball's movement. We also asked participants whether the sound was that of a departure or an arrival. Figu 4 show the illustration of this user study.

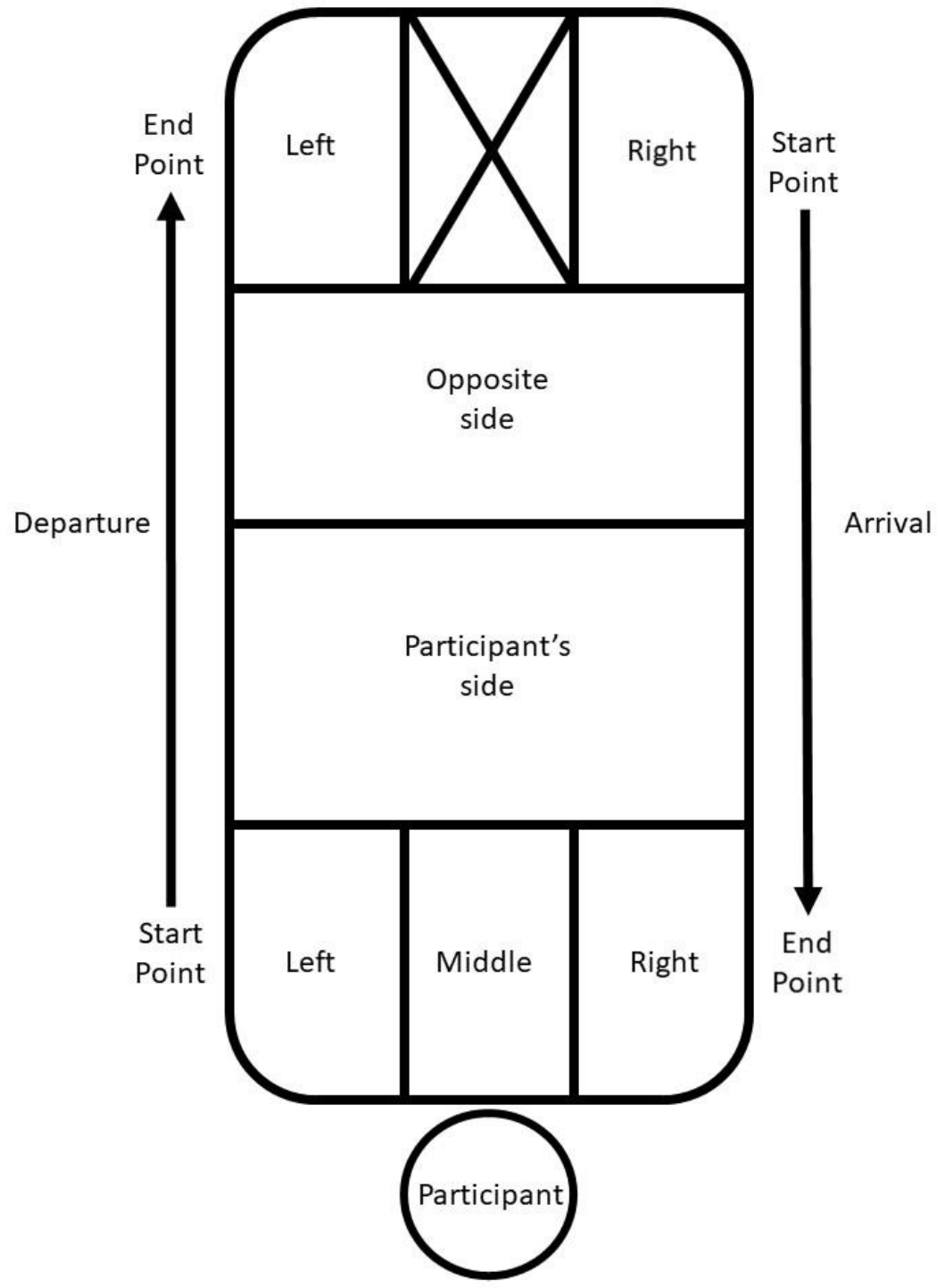

Fig 4. Illustration of User Study 1.

For quantitative evaluation, we measured the participant's accuracy in detecting the correct route. We also calculated the accuracy rate for each side (participant side vs. opposite side), for each route (departure vs. arrival), and for start points vs. end points.

*5.1.3 Procedure.* Each participant signed a consent form and answered demographic questionnaires. All participants wore eye patches which covered both eyes during the user study. The participants stood at the showdown table as shown in fig 4. Each participant had training in order to become familiar with the sound of the showdown ball. After training, the participants answered eighteen questions about the start and end of the ball moving away from participant using their hearing only. The order of the questions was changed via a Latin square. After answering the questions, participants had a five-minute break. Then participants answered another eighteen questions about the start and end of ball moving towards the participant also using their hearing only.

*5.1.4 Results.* We analysed the mean of overall accuracy rate, the accuracy rate for each side (the participant's side and the opponent's side), for each direction (departure, arrival). All accuracy rates were analysed by a Mann-Whitney U test because the Shapiro-Wilk normality test showed that neither both groups of each answer rate showed normal distribution. The overall accuracy rate showed an average of above 91% for both participant groups. There was no significant difference between people with visual impairments and people with able-body in any of the trial questions. Fig 5 shows all correct recognition rates of ball's position and direction by the two participating groups.

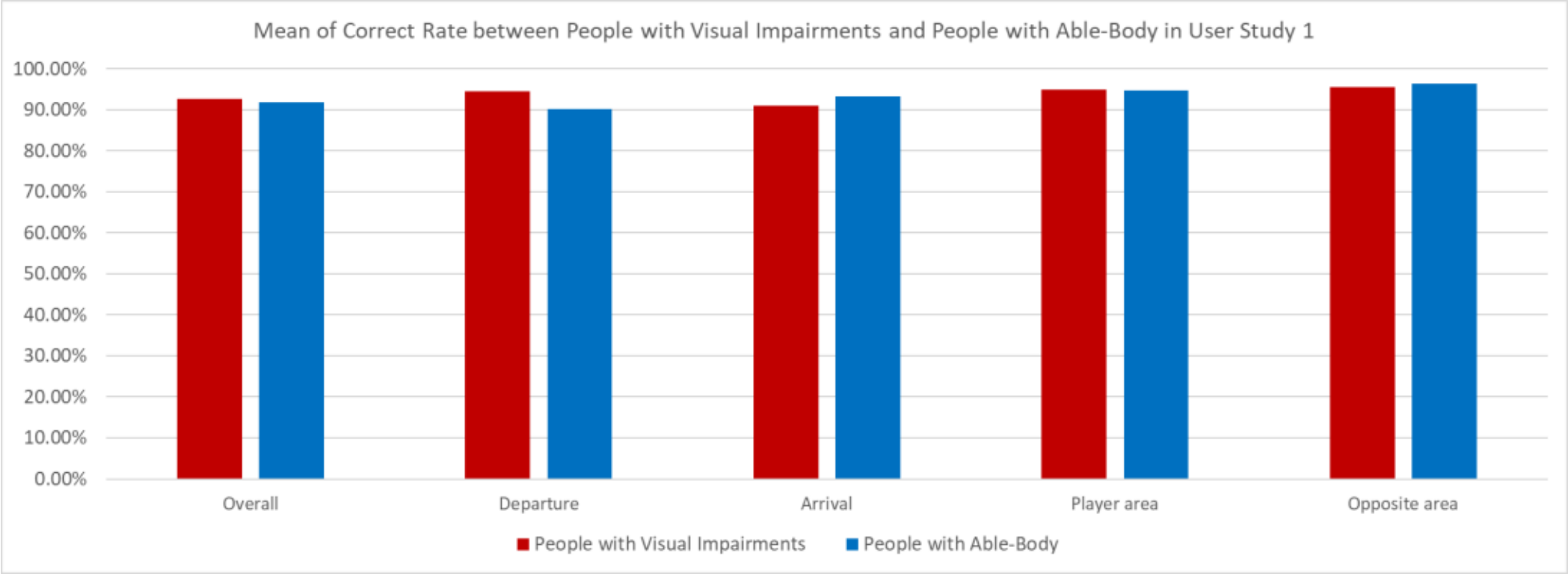

Fig 5. Comparison of correct recognition rates of ball's position and direction by two participant groups.

## 5.2 User Study 2: Validation of the PVA mode

In the second user study, we tested the PVA mode for people with visual impairments. The participants played a best-of-three series of 12-point games. We expected the player to hit the showdown ball if the player detected its position. The player intended to move the ball to the opponent's goal to win. Furthermore, in comparing people with visual impairments and people with able-body, related studies show that people with visual impairments are better than people with able-body at performing auditory spatial tasks [48]. Although not all people with visual impairments have superior skill in auditory spatial perception [60], there are studies that show that people with visual impairments are generally more sensitive to small changes in auditory tone than people with able-body [11, 61]. Because HRTF changes the timbre of sound based on the virtual physical system, we expected people with visual impairments to be better than people with able-body at tracking the ball's movement and speed due to their heightened auditory sensitivity. Thus, we expected they would hit more balls than participants with able-body they are more sensitive to auditory changes. Farther more, we expect people with visual impairments can hit the into opponent's goal hole based on high auditory sensitivity. Therefore, we posited the following hypothesis for verification.

H2. In PVA (Player vs. Agent) mode, people with visual impairments have a higher probability of hitting the approaching ball into the opposing *area* and a higher probability of hitting the ball into the opponent's *goal* than people with able-body (but blind-folded).

We also want to know that the PVA mode can give exertion effect as other exergame present. To confirm this, we record heart rate of people with visual impairments and analyse as maximum heart rate percentage. The related study showed that the interval (64% to 76%) of maximum heart rate percentage shows that the exercise is moderate-intensity [14]. We calculated the maximum heart rate percentage by the following formula: (maximum heart rate/220-age) [10, 34]. This method is a well-known formula for researching VR exergame [3, 26, 35, 70].

*5.2.1 Participants and Apparatus.* 13 participants with able-body and 13 people with visual impairments participated in this user study. The age range of the participants with able-body was 20~27 (M=22.23). They were all university students with normal or corrected to normal vision and with normal hearing ability. In addition, all able-body participants had previous experience in VR environments. We used the PVA mode from our developed system so the participants competed with an AI agent. Please see a detailed description of PVA mode in the implementation section.

*5.2.2 Tasks and measurement.* The task for participants was to win two games against the AI agent. Each participant received cash rewards of four dollars for winning the two games. We calculated systematic measurements including scores for quantitative analysis (see Table 2). To measure the heart rate of people with visual impairments, we used Fitbit charge3, which is a smart watch that measures heartrate via LEDs and light sensitive photodiodes [16] and calculated the maximum heart rate percentage.

Table 2. The quantitative measurement which was calculated by the game's system in user study 2.

| Collected data | Definition |
|---|---|
| Match Win | Number of participants who win two games first |
| Game Win | Number of games which participants get 12 points first |
| Shots on Target | Number of balls that went near the opponent's goal |
| Hit | Number of hits on the ball by a participant |
| Left Hit | Number of hits from the left by a participant |
| Middle Hit | Number of hits from the middle ball by a participant |
| Right Hit | Number of hits from the right ball by a participant |
| Match Win Rate | Number of participants who won the match/Number of matchers |
| Game Win Rate | Number of games participants won/Number of games |
| Shots on Target Rate | Number of shots on target/Number of hits |
| Hit Rate | Number of Hits/(Number of hits + Number of misses) |
| Left Hit Rate | Number of Left Hits/(Number of left hits + Number of left misses) |
| Middle Hit Rate | Number of Middle Hits/(Number of Middle hits + Number of misses) |
| Right Hit Rate | Number of Right Hits/(Number of Right hits + Number of misses) |

*5.2.3 Procedure.* Each participant signed a consent form and answered questionnaires on demographics. All participants wore eye patches during the user study. The participants had training time to become familiar with the virtual showdown game. The participants played a best-of-three series of 12-point games against the AI agent. The participants played three games unless they won the first two games in which case they only played two games.

*5.2.4 Results.* To analyse the results, we calculated the average value of each variable for each participant because they did not all play the same number of games. We used an independent t-test and a Mann-Whitney U test based on the results of a Schapiro-Wilk normality test.

For an overall result, ten out of thirteen people with visual impairments and nine out of thirteen people with able-body won the whole series in PVA mode. The group of people with visual impairments won 21 games out 30 games, which is a 70% winning rate. The people with able-body (albeit wearing eye patches) won 21 games out 29 games as a group, which is a 72.41% wining rate. Because each participant played a different number of games, all data except Match Win Rate and Game Win Rate are based on the mean of each participant. We did not found any significant difference in quantitative data except on “shots on target”. We found that people with visual impairments had a significantly higher rate of “shots on target” than people with able-body (U=20, p=0.000). The results show that hypothesis 2 is partially supported because we didn’t find that people with visual impairments had significantly higher rate of hit than people with able-body. Fig 6 shows the values of the collected data.

The results of heart rate monitoring show the mean maximum heart rate is 112.15 and the mean max heart rate percentage is 69.97%. This result showed that the PVA mode of the game induces moderately intense exercise because the max heart rate percentage interval of moderate-intensity exercise is 64% to 76% [14]. Table 5 shows the details of heart rate monitoring in user study 2.

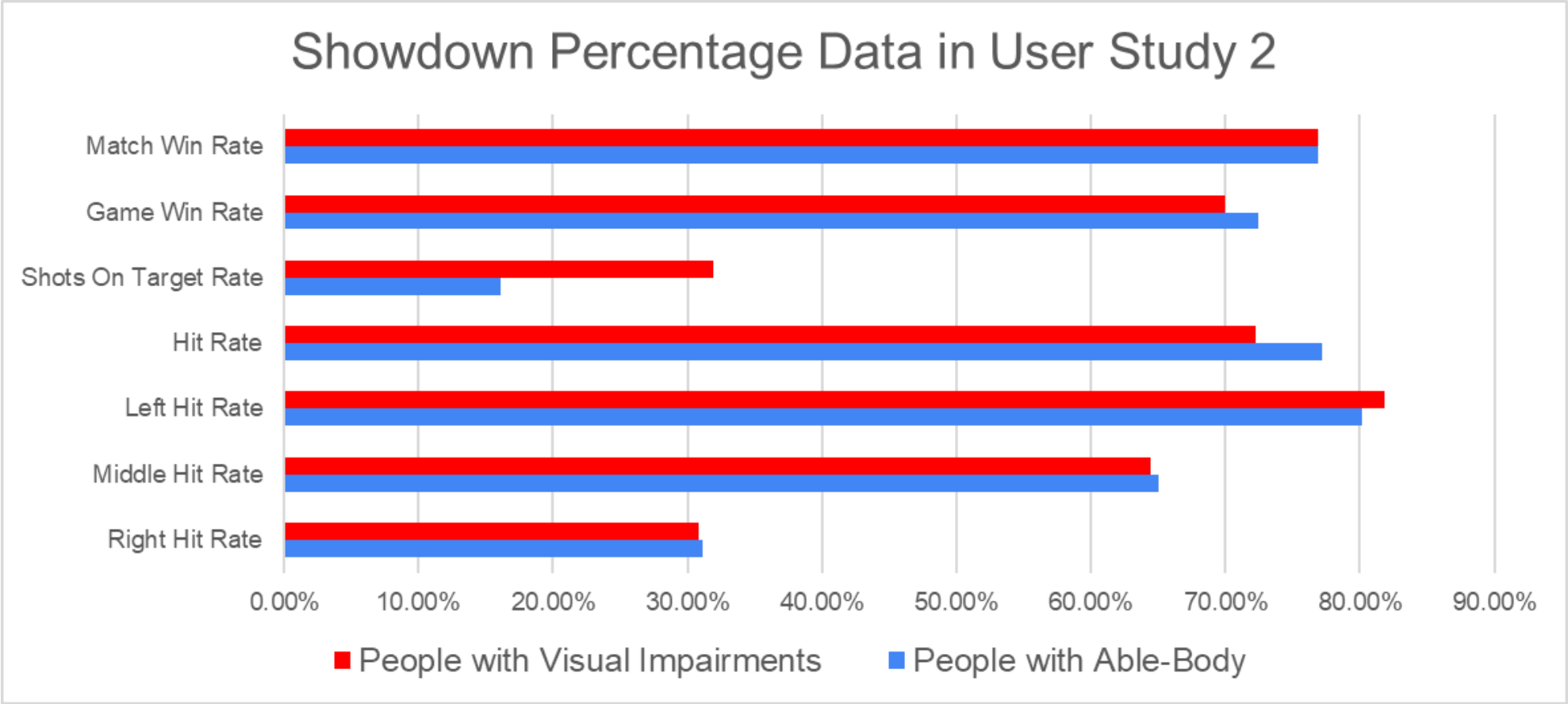

Fig 6. Result of game data of two participant groups in User Study 2

### 5.3 User Study 3: Validation of PVP mode

In the third user study, we tested the PVP mode by comparing the two groups in the same game. One interesting feature of the PVP mode is that the ball travels much faster and less predictably than in the PVA mode. According to related work [11, 61], people with visual impairments are better at perceiving distance via audio than people with able-body. In addition, people with visual impairments are better at moving the ball into the goal than participants with able-body; this is confirmed by our user study 2 results. Therefore, we expect the people with visual impairments to win more games when competing against people with able-body. However, it is uncertain whether the auditory perception of people with visual impairments will be better than that of people with able-body in the context of sports where the position of sounds changes rapidly.

H3. In PVP (Player vs. Player) mode, people with visual impairments will defeat people with able-body more of than they will be defeated by them.

We also want to know that the PVP mode can give exertion effect as other exergame present. To confirm this, we record heart rate of people with visual impairments and analyse as maximum heart rate percentage as same as user study 2. Farther more, we compared heart rate data between user study 2(PVA mode) and user study 3(PVP mode).

*5.2.1 Participants and Apparatus.* For this user study, we recruited an additional thirteen able-bodied participants. The age range of abled-bodied participants was 21~56 (M=37.92). Seven were university students and seven were assistants of people with visual impairments. Six people had previous VR experience and twelve people were right-handed.

We used the virtual showdown software which was implemented as reported in the implementation section. In addition, we used the PVP mode in which one participant can compete with another human participant. There is a detailed description of PVP mode in the implementation section. To measure the heart rate of people with visual impairments, we used Fitbit charge3, which is a smart watch that measures heartrate via LEDs and light sensitive photodiodes [16] as same as user study 2.

*5.2.2 Tasks and measurement.* Participants aimed to win two games against their opponent. Participants who won two games first received a cash reward of four dollars. In addition to the same measurement as used in user study 2, we collected three additional kinds of data. Table 3 shows the additional data we collected in user study 3. All data was automatically collected and calculated by the computer log system.

Table 3. Additional quantitative measurements applied in user study 3.

| Collected data | Definition |
|---|---|
| Rally | Number of balls returned when a participant hits the ball |
| Number of Ball sent | Number of balls sent from the participant's area to the opponent's area |
| Number of approaching Balls | Number of balls sent from the opponent's area to participant's area |

For this user study, we also used a questionnaire to evaluate the participants' perceived VR experiences. The user study questionnaire was based on the Stanney et al. [55]'s questionnaire. We used a 5-point Likert scale, where the higher score indicates a more positive response. For the easy understanding of people with visual impairments, we modified the questions to suit the showdown VR context. Table 4 shows the questionnaire we used in this user study. We also interviewed each participant to get detailed information about their VR showdown game experience. After completing the questionnaire, participants answered questions about the ease of use of the VR system and the appropriateness of the audio feedback. The interview questions were open-ended in order to encourage the sharing of a wider range of opinions about the participants' VR experiences.

*5.2.3 Procedure.* Each participant signed a consent form and answered demographic questionnaires. All participants wore eye patches on both eyes during the user study. The participants had training time to become familiar with the virtual showdown game. Participants with able-body took more training time to become familiar with the game system. This was intended to reduce a learning effect gap between the two participant groups. They played a best-of-three series of 12-point games against each other. They played three games except when either participant won the first and second games. After playing the games, the participants answered the questionnaire and were interviewed by a researcher.

*5.2.4 Results.* We calculated the average value of each variable based on each participant because the participants did not necessarily play the same number of games. For the statistical method, we used an independent t-test and a Mann-Whitney U test based on the results of a Schapiro-Wilk normality test.

For overall results, 11 out of 13 people with visual impairments won against their opponents with able-body. From the analysis of each game, the people with visual impairments won 23 of 34 games, a 67.6% win rate. From a comparison of the goal count between participant groups, there was no significant difference except in the left hit rate. The mean of the left hit rate between the groups was 90.12% and 55.97% in favour of the group with visual impairments, a strongly significant difference (F (13.829) = 4.520, p=0.000). Fig 7 shows the percentage data between two participant groups. Fig 8 shows the mean of each collected type of data.

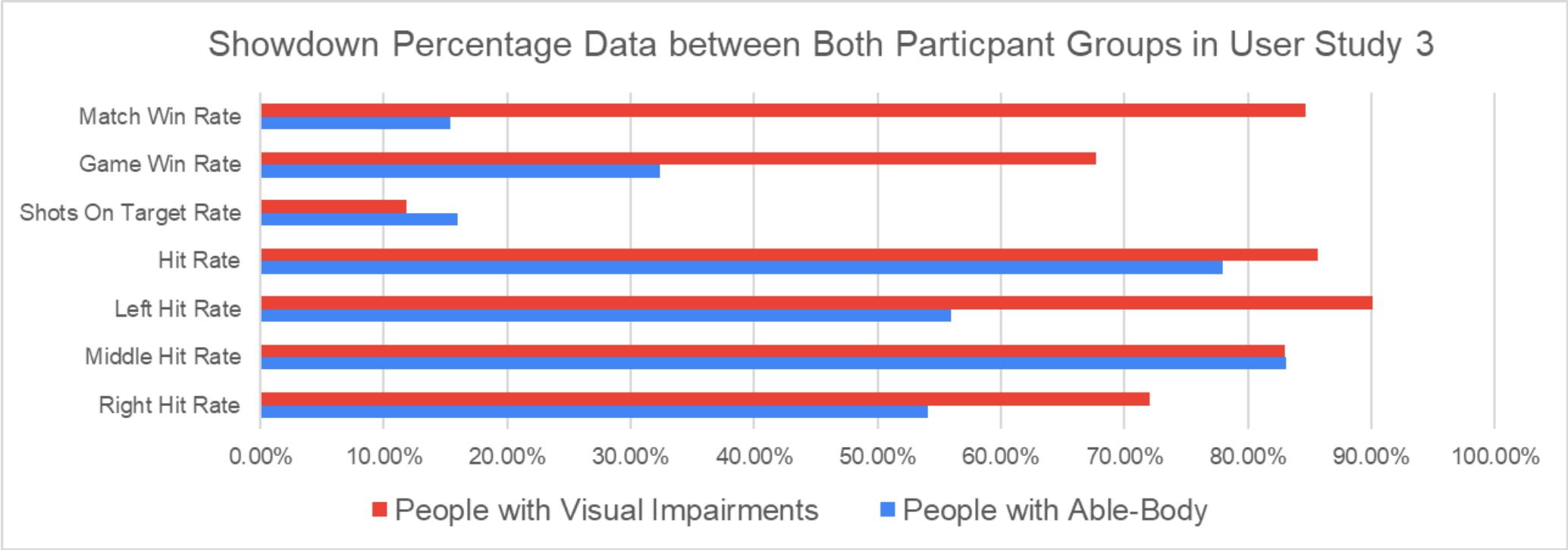

Fig 7. Rate of each percentage data, collected between the two participant groups.

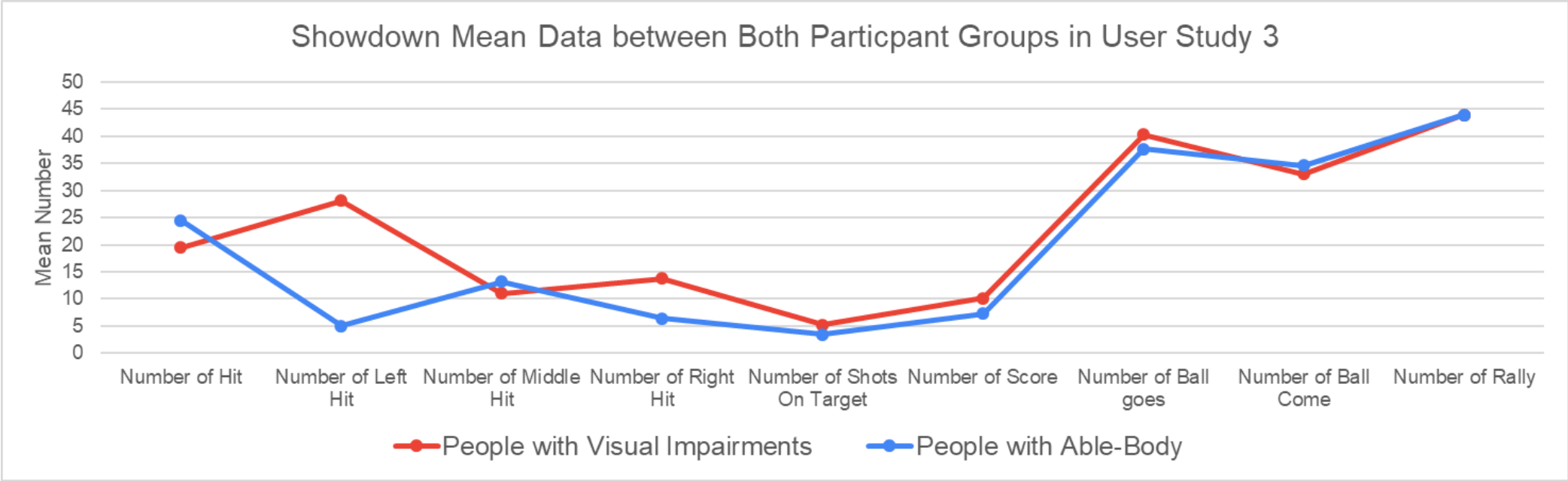

Fig 8. Result of each mean data we collected between the two participant groups.

In the questionnaire, we found a significant difference between the groups for two of the questions. All questionnaire data were analysed with a Mann-Whitney U test because all data was abnormal according to the Schapiro-Wilk normality test. The responses to the question about the integration of audio feedback, (U=39.5, p<0.01) and overwhelming physical fatigue (U=42, p<0.05) showed significant differences. Table 4 shows the questionnaire result, which shows significant differences in question 11 and 19.

Table 4. The mean and standard error of questionnaire between two participant groups in user study 3. (*p<0.01, **p<0.05)

| Questionnaire | | People with visual impairments | People with able-body |
|---|---|---|---|
| 1. | The position of the virtual showdown ball was recognizable [18]. | 4.08 (0.329) | 3.62 (0.266) |
| 2. | The device helped me to notice the position of the virtual showdown ball [6]. | 4.69 (0.175) | 4.38 (0.241) |
| 3. | The VR device presents information that helps me to know where I want to send the virtual showdown ball [64]. | 4.62 (0.180) | 3.85 (0.337) |
| 4. | The audio feedback was helpful to get information about the direction of the showdown ball [66]. | 4.85 (0.154) | 4.85 (0.104) |
| 5. | It was easy to change or rearrange the position of the virtual showdown ball [18]. | 4.15 (0.355) | 3.77 (0.303) |
| 6. | The ball movement was responded naturally to the controller’s movements [54]. | 4.54 (0.215) | 4.23 (0.231) |
| 7. | Ball movement in VR was free enough [18]. | 4.77 (0.122) | 4.69 (0.133) |
| 8. | It was easy to use and control the input device (Controller) [18]. | 4.54 (0.215) | 4.85 (0.104) |
| 9. | The input device (Controller) was easy to use so that I understood the spatial state of the virtual showdown ball and myself [51]. | 4.46 (0.291) | 4.31 (0.237) |
| 10. | The feedback (feeling) after using the input device (Controller) was well presented to me [25]. | 4.77 (0.122) | 4.69 (0.133) |
| 11. | The auditory output was perfectly integrated with my activity in VR [9].* | 4.77 (0.166) | 4.00 (0.226) |
| 12. | The auditory output had no lag or inconvenient characteristics [47]. | 4.54 (0.243) | 4.62 (0.140) |
| 13. | The auditory output was meaningful, timely, and useful [66]. | 4.69 (0.175) | 4.54 (0.183) |
| 14. | I felt that I was part of VR while using the VR device [18]. | 4.92 (0.077) | 4.54 (0.243) |
| 15. | I was able to use the controller naturally and to react to the virtual environment fluidly [65]. | 4.85 (0.104) | 4.54 (0.144) |
| 16. | I thought that the VR experience in this user study was similar to what I had experienced before [65]. | 4.46 (0.312) | 4.08 (0.211) |
| 17. | I did not feel significant discomfort while using the device [9]. | 4.62 (0.18) | 4.54 (0.268) |
| 18. | Fatigue did not increase with time while I was using the device [9]. | 4.46 (0.215) | 3.77 (0.323) |
| 19. | Overwhelming physical fatigue was not evoked while using device [72].** | 4.62 (0.213) | 3.85 (0.249) |
| 20. | I did not feel overwhelming discomfort while in VR [20]. | 4.69 (0.237) | 4.69 (0.133) |
| 21. | I did not feel any headache while in VR [20]. | 5.00 (0.00) | 4.77 (0.166) |
| 22. | I did not sweat a lot [20]. | 4.46 (0.215) | 4.92 (0.077) |
| 23. | I did not feel nauseous [20]. | 5.00 (0.00) | 4.77 (0.166) |
| 24. | I want to use this VR system again [55]. | 4.62 (0.213) | 4.54 (0.268) |
| 25. | I did not feel any headache or nausea after using the VR device [20]. | 5.00 (0.00) | 4.69 (0.175) |
| 26. | I did not experience vertigo (loss of orientation to vertical upright) after using the VR device [20]. | 5.00 (0.00) | 4.69 (0.175) |

From the result of the heart rate, the mean of the maximum heart was 121.31 and the mean percentage of max heart rate was 75.45%. This percentage of max heart rate shows that the PVP mode also evoked a moderate-intensity exercise level (64%~76%) [10]. We compared the mean maximum heart rate and the mean max heart rate percentage from each user study. We used the paired t-test because both heart rate data are normally distributed by the Shapiro-Wilk normality test. The test results revealed no significant difference between the two user study studies', max heart rate and max heart rate percentage. Nevertheless, heart rate data show a tendency for PVP mode to induce a higher heart rate than PVA mode. The table 5 show the mean and standard error of maximum heart rate and the max heart rate percentage between two user studies.

Table 5. The mean and standard error of maximum heart rate and the max heart rate percentage.

| Played Mode | PVA (Player vs. AI agent) mode (User Study 2) | PVP (Player vs. Player) mode (User Study 3) |
|---|---|---|
| Max heart rate | 112.15(4.25) | 121.31(3.96) |
| Max heart rate percentage | 69.97% (3.09%) | 75.45% (2.48%) |

## 6 DISCUSSION

Through three user studies, we tested our hypotheses and found meaningful results that inform regarding the accessibility potential of people with visual impairments to VR. These user studies confirmed H1 by achieving over 91% for overall accuracy in both participant groups. Another similar study, which tested perceived distance to the audio source using HRTF with visually impaired participants showed an average of about 80% [60]. Like other previous studies, our first user study showed that the HRTF offers great performance delivering the perception of audio distance and even changes in the position of the audio source in real time. Furthermore, overall accuracy is almost the same in both participant groups. User study results also showed that HRTF could give adequate spatial information through sound to people with able-body just as with people with visual impairments. This result supports the conclusion of the previous study, that HRTF can deliver accurate spatial information to people with able-body [29]. Also, this result supports why many VR for people visual impairments used HRTF [22, 27, 28, 53, 73] .

From the result of second user study, Hypothesis 2 was supported in that people with visual impairments are significantly better at hitting the ball to the opponent's goal. This can be interpreted to mean that people with visual impairments are better than people with able-body at predicting the trajectory of the ball. In addition, the accuracy rate for the left side hit was higher than that of the middle and right side hit. This reflects the fact that all people with visual impairments were right handed making it easier for them to hit the ball from the left side, considering that they hit the ball by pushing the controller like an air hockey puck.

The results of the third user study show that 11 of 13 people with visual impairments won the match (the whole set of games) which supports Hypothesis 3. As similar results in user study 2 show, people with visual impairments hit the ball with 70% accuracy and the accuracy of hitting the ball from the left was higher than hitting the middle and right ball. People with visual impairments achieved significantly higher left ball hit rate accuracy than people with able-body. Even though twelve people with able-body was right-handed, still people with visual impairments can detect and hit the left ball more easily than they can. From the questionnaire, we found that people with visual impairments gave a high score for audio integration. They also said that they were not overwhelmed by physical fatigue in the interview. By contrast, participants with able-body said that they felt tired. We assume that the people with able-body felt physical fatigue because they were not used to dealing with situations in which they could not see. Also, people with visual impairments were more sensitive to auditory changes so that more detailed auditory distance information would have been recognized using less cognitive resources than the people with able-body [11, 61].

After the user study, in the interview, we found that seven out of thirteen people with visual impairments responded saying that they could make a mental map of which direction the ball will move when they hit the ball. This statement supports related studies and it could support our user study result that people with visual impairments showed better performance when hitting the ball into the opponent's goal than people with able-body. In addition, we found that all eleven people with visual impairments said their previous original Showdown game experience helped them play our VR showdown game. People with visual impairments who had experience in regional Showdown championships said, “The process of recognizing the position and sound of a moving ball and hitting the ball to the opponent's goal is almost the same as in the real Showdown game.” With this comment, we expect that our VR system can produce a very similar experience to the original Showdown game.

One people with visual impairments said, “While playing with another person, I enjoyed the game more than playing with AI because the moving direction of the ball was less predictable every time.” We confirmed that the PVP mode is more enjoyable because the ball’s movement changes unpredictably by hitting in response to the unpredictable behaviour of the un-programmed opponent, unlike in the PVA mode. This result also supports Federoff et al.’s work, which states that game AI should be unpredictable [8].

We developed a holding function for players to move the ball easily into the position they wanted. We expected that holding would add to the available strategies, help the player to play more speedily and thereby add the experience. In the user study, all people with visual impairments answered that they understood how holding worked in the training sessions. For this reason, we initially expected that participants who used holding well might win the game. In user study 3, six of thirteen people with visual impairments and ten of thirteen participants with able-body used holding function. We found that this interaction was natural to some participants but not all participants. In further study, we suggest implementing another holding mechanism which is used in the original Showdown. In a real Showdown situation, players stop the ball's movement by pushing the ball to the sidewall using the showdown racket.

For motion tracking, people with visual impairments are more sensitive to the sound and whether it matches well with their own action. This result shows that people with visual impairments can play the exergame with more precise movements in 3D virtual environments using auditory distance perception over gestures detected exergames [37-40].

The interviews revealed that the preferred environment is that in which real competitors move the ball in a complicated and unpredictable manner rather than the more predictable moves of a programmed AI opponent. In other words, playing with human opponents is more enjoyable than competing with a computer as the related exergame studies for people with visual impairments show [37, 38].

Finally, twelve out of 13 people with visual impairments answered "yes" when asked if they would like to play our VR showdown game at home. As a possible further study, we will transplant the game into a mobile platform so that the people with visual impairments can enjoy playing the game against other people with visual impairments or people with able-body in any place.

Due to the limited recruitment availability of the people with visual impairments, most of the participants with visual impairments were elderly, and they participated in all three studies. Thus, in the third user study in which they competed with able-bodied people, there may be learning effects for people with visual impairments. On the other hand, participants with able-body may have had physical advantages because they were much younger than the participants with visual impairments. Despite these potential disadvantages for the elderly, the results show that the people with visual impairments performed better than the abled-bodied people in virtual showdown game. For future work, the third user study of the showdown match needs to be re-examined with participants of the similar ages and similar knowledge for showdown to overcome possible learning effects.

## 7 DESIGN GUIDELINE

We propose a design guideline for a collaborative VR system for people with visual impairments and people with able-body. The premise and the general design principle is to "level the playing field" by providing equalizing conditions, which, in our case meant providing the same feedback to both people with visual impairments and people with able-body. Since our game only gives audio and haptic feedback, the visual feedback was removed from all participants. We therefore offer the following suggestions:

*HRTF is not appropriate to produce realistic spatial sound from a distance in virtual reality*. HRTF is a well-known spatial acoustics method in VR and it is used in many VR systems for people with visual impairments [22, 27, 28, 53, 73]. However, HRTF is not a panacea that can produce all the spatial sounds of a virtual environment. As the virtual sound source gets further away, it is difficult to feel the location of objects because it is difficult to create an acoustic parallax that offers a sufficient angle difference between the virtual sound source and each ear. To implement acoustic parallax, the prerequisite is that the virtual audio source must be close enough to the listener's head so there is a significant angular difference between the two ears and the audio source [71]. As the virtual source of sound recedes into the virtual distance, the angle required to differentiate between left and right aspects of the source approaches zero, i.e., less and less distinction between left and right is detectable. In our experiment, we adjusted the size of the showdown table appropriately to sustain the HRTF effects of moving ball over the whole playing area. VR developers should be aware that HRTF is not suitable for applications that require accurate spatial information from an audio sources which is far away.

*VR interactions don't have to be exactly the same as real-world interactions in the VR experience of people with visual impairments.* Many VR systems for people with visual impairments have tried to implement the same interaction and feedback as in the real world. For example, Zhao et al. and Siu et al. developed a cane-shaped VR controller that cannot move across solid virtual objects [53, 73]. However, artificial VR interactions can enhance the VR experience over realistic VR interactions [30]. For example, the Go-Go technique allows the user to grasp a distant virtual object by stretching the virtual hand [44]. Thus, the player does not have to move to catch distant objects. In our system, we have developed an artificial interaction, the holding function. The holding feature allows the player to move the ball as intended, following the controller while holding the trigger button. This method avoids physical engine errors that cause the player to move the ball in unintended ways, and so people with visual impairments could make fewer mistakes in controlling the ball. Through training in the second and third user studies, participants said that the holding function was natural and easy to understand in interviews after the studies. They mentioned that the holding function allows them to use various strategies to win the game. This artificial feature allowed people with visual impairments to have a satisfying VR experience i.e., a real experience within their parameters.

*The collaborative exergame may produce an improved exercise experience over single-player exergames.* We have developed a PVA mode (single player mode) and a PVP mode (multiplayer mode) for our showdown VR game. We also conducted a user study of both modes for both people with visual impairments and people with able-body. Two user studies have shown that the percentage of maximum heart rate for people with visual impairments is relatively higher in the PVP mode than in the PVA mode. Although the difference in heart rates between the two studies is not significant, the PVP mode tends to induce more benefits for people with visual impairments. We also found this trend in the surveys after the second and the third user study. The ratings to questions about the immersion and usefulness of the controller tend to be high in the third user study. These results support the view that multiplayer exergames could provide a better exercise experience than single-player exergame. In a similar study, Kors et al. showed that group players play exergames longer and more often than a single player games [19]. The collaborative exergame not only boosts the exertion effect but also improves the social relationship. For example, Saksono et al. suggested a collaborative exergame for families in which parents and children can exercise together and complete the game's quests together [49]. Ren et al. developed the social exergame for office workers which evokes social interaction to complete collaborative fitness tasks [46]. Therefore, through this collaborative VR exergame, people with visual impairments can achieve enhanced exercise effects, general satisfaction, and social-presence implications despite the lack of vision and visual communication cues.

## 8 CONCLUSIONS

For this study, we developed a collaborative VR showdown game, which is accessible to people with visual impairments using Vive and HRTF. For this game, we used HRTF so that people with visual impairments could detect a rapidly moving showdown ball using auditory feedback only. In addition, we used Vive Pro to track the position and movements of the head and hand of participants in real time. We tested our developed system on people with visual impairments and also on people with able-body. We found that people with visual impairments can detect the position of the showdown ball with 90% accuracy. We also found that people with visual impairments are better at hitting the ball to the opponent's goal than people with able-body. Lastly, by monitoring heart rates, we confirmed that people with visual impairments can experience the effects of exercise with a similar intensity to that obtained through playing table tennis. This research extends the VR exergame field of knowledge for people with visual impairments and offers promising directions for further research regarding accessible VR for people with visual impairments.

## ACKNOWLEDGMENTS